\documentstyle[12pt]{article}
\begin{document}

\font\fortssbx=cmssbx10 scaled \magstep2
\hbox to \hsize{
\hbox{\fortssbx University of Wisconsin - Madison}
\hfill$\vcenter{\hbox{\bf MADPH-96-931}
                \hbox{\bf hep-ph/9602279}
                \hbox{February 1996}}$ }

\vspace*{20mm}

\baselineskip22pt
\begin{center}
\Large \bf 
Radiative corrections to sfermion mass splittings
\end{center}
\vspace{10mm}

\begin{center}
\large Youichi Yamada
\end{center}
\vspace{0mm}

\begin{center}
\begin{tabular}{l}
{\it Physics Department, University of Wisconsin, Madison, WI 53706, USA}
\end{tabular}
\end{center}

\vspace{20mm}
\begin{center} 
\large Abstract
\end{center}
\begin{center}
\begin{minipage}{13cm}
\baselineskip=20pt
\noindent
We study the one-loop radiative corrections to the SU(2) breaking mass 
splittings between sfermions in SU(2) doublets 
in the Minimal Supersymmetric Standard Model. 
At tree-level, the differences of mass squared 
$m_{\tilde{f}_{1L}}^2-m_{\tilde{f}_{2L}}^2$ of SU(2) doublet sfermions 
$(\tilde{f}_1, \tilde{f}_2)_L$ in the first two generations 
are determined by $\tan\beta$, and 
are universal to sleptons and squarks. 
The radiative correction, however, breaks this relation. The typical 
deviation from the universality between sleptons and squarks 
is within $\pm0.05$ in terms of the ``effective $\cos2\beta$''. 
We also study the SUSY parameter dependence of the deviation. 
For very heavy sfermions, the relative corrections become large. 
\end{minipage}
\end{center}
\newpage

\baselineskip=22pt
\normalsize
\setcounter{footnote}{0}

\section{Introduction}
\hspace{\parindent} 
In the Minimal Supersymmetric (SUSY) Standard Model (MSSM) \cite{mssm}, 
the tree-level masses of left-handed sfermions $\tilde{f}_L$ in SU(2) doublets 
$\tilde{F}=(\tilde{f}_1, \tilde{f}_2)_L$ in the first two generations 
are given in terms of SU(2) invariant masses $M_{\tilde{F}}$, the ratio of 
vacuum expectation values of two Higgs scalars $\tan\beta=v_u/v_d$, 
isospin $I_{3f_L}$, charge $Q_{f_L}=I_{3f_L}+Y_{f_L}$, $m_Z$ and 
$s_W\equiv\sin\theta_W$, ignoring masses of their fermionic 
superpartners $f$. The explicit form of their masses is 
\begin{equation}
m_{\tilde{f}_L}^2=M_{\tilde{F}}^2+(I_{3f}-s_W^2Q_f)m_Z^2\cos2\beta . 
\label{eq1}
\end{equation}
The difference of mass squared between $\tilde{f}_{1L}$ and 
$\tilde{f}_{2L}$ is then 
\begin{equation}
m_{\tilde{f}_{1L}}^2-m_{\tilde{f}_{2L}}^2=m_W^2\cos 2\beta. \label{eq2}
\end{equation}
Eq.(\ref{eq2}) is independent of 
flavors (mass, color, hypercharge) of sfermions and also of any 
unification conditions beyond the MSSM such as the unification of 
sfermion masses in minimal supergravity. 
As a special case, a mass sum rule between squarks and sleptons 
\begin{equation}
m_{\tilde{\nu}}^2-m_{\tilde{\ell}_L}^2
=m_{\tilde{u}_L}^2-m_{\tilde{d}_L}^2(=m_W^2\cos2\beta), \label{eq3}
\end{equation}
follows from eq.(\ref{eq2}). 

The reason for the relation (\ref{eq2}) is that 
the mass splitting between $\tilde{f}_{1L}$ and $\tilde{f}_{2L}$ is generated 
by the D-term four-scalar interaction $\lambda(F^*\tau^aF)(H_i^*\tau^aH_i)$ 
($i=(d,u)$, $\tau^a$: generators of SU(2)) which is related to 
the SU(2) gauge 
vector interaction by the SUSY Ward identity. 
The relations (\ref{eq2},\ref{eq3}) then provide a detailed test of 
the MSSM \cite{martin,feng} and also a method for fixing 
$\tan\beta$ \cite{ay,brig}. 

However, due to the SUSY violation, the relation (\ref{eq2}) between 
sfermion masses, $m_W$ and $\tan\beta$ is modified 
by radiative corrections. 
In this paper, we present analytic and numerical results 
of one-loop corrections to mass splittings (\ref{eq2}) 
of SU(2) doublet sfermions in the first two generations (i.e. SUSY 
partners of massless quarks and leptons). 
The breaking of the relation (\ref{eq2}) was briefly discussed in 
ref.\cite{ay} for 
$(\tilde{f}_1,\tilde{f}_2)_L=(\tilde{\nu},\tilde{e}_L)$ case. 
Here we mainly discuss the breaking of the mass relation (\ref{eq3}) 
between sleptons and squarks. 
It is shown that the deviations 
from the relation (\ref{eq3}) are typically within $\pm0.05$ in terms of 
``effective $\cos2\beta$'', which is defined in section 3. 
The relative corrections to the tree-level results (\ref{eq2}, \ref{eq3}), 
however, become large for very heavy sfermions. 
The measurement of the violation of the relation (\ref{eq3}) 
would therefore be important to know the nature of the MSSM beyond the 
tree-level.

This paper is organized as follows. In section 2, we present the analytic 
form of the one-loop corrected mass splitting for sfermions. The 
renormalization condition is also briefly discussed. In section 3, we 
show the numerical results of the radiative corrections to 
eqs.(\ref{eq2}, \ref{eq3}) and their dependence on various SUSY parameters. 
Section 4 is devoted for conclusion. 

\section{One-loop corrections to sfermion masses}
\hspace{\parindent} 
In this paper, we ignore the mixing of sfermions in different generations. 
Then 
the one-loop corrected mass of a sfermion $\tilde{f}_L$ is expressed 
as 
\begin{equation}
m_{\tilde{f}_L}^2({\rm pole})=\hat{m}_{\tilde{f}_L}^2
-{\rm Re}\Pi_{\tilde{f}_L}(q^2=m_{\tilde{f}}^2)
+2m_Z^2(I_{3f}-s_W^2Q_f)(\cos^2\beta\frac{\Delta v_d}{v_d}
-\sin^2\beta\frac{\Delta v_u}{v_u}), \label{eq4}
\end{equation}
where $\hat{m}_{\tilde{f}_L}$ denotes the tree-level mass (\ref{eq1}) 
in terms of 
tree-level parameters ($\hat{M}_{\tilde{F}}^2$, $\hat{m}_Z$, $\hat{s}_W$, 
$\tan\hat{\beta}$):
\begin{equation} 
\hat{m}_{\tilde{f}_L}^2=\hat{M}_{\tilde{F}}^2+(I_{3f}-\hat{s}_W^2Q_f)
\hat{m}_Z^2\cos 2\hat{\beta} .
\end{equation} 
$\Pi_{\tilde{f}_L}(q^2)$ is the two-point $\tilde{f}_L^*\tilde{f}_L$ function. 
$\Delta v_{d,u}$ denotes the shift of vacuum 
expectation values from the tree-level values, namely sum of 1-loop tadpole 
contributions $\Delta T_{d,u}^{(1)}$ and their counterterms 
$\Delta T_{d,u}^{(1)CT}$. 
We comment on the renormalization of $\Delta v_{d,u}$ later in this section. 

We use the dimensional reduction \cite{dr} for regularization. 
The contributions of the $\epsilon$-scalar masses \cite{epsilon} are 
neglected here since they are common to $m_{\tilde{f}_1}$ 
and $m_{\tilde{f}_2}$ and cancel out in the mass difference (\ref{eq2}). 

We calculate $\Pi_{\tilde{f}_L}(q^2)$ in terms of 
the 't Hooft-Veltman functions \cite{pv} $A$, $B_{0,1}$, 
using definitions given in Ref.\cite{hhkm}. 
We adopt the 't Hooft-Feynman gauge for convenience. 
As for couplings of SUSY particles, we basically follow the 
notations in Ref.\cite{gh}. 
The explicit form of $\Pi_{\tilde{f}_L}(q^2)$ is then 
(henceforth we omit the subscript $L$ for $\tilde{f}$, $\tilde{f}_{1,2}$.): 
\begin{eqnarray}
\Pi_{\tilde{f}}(q^2)&=&\Pi_{\tilde{f}}^{g,\tilde{g}}(q^2)
+\Pi_{\tilde{f}}^{\gamma Z,\tilde{\chi}^0}(q^2)
+\Pi_{\tilde{f}}^{W,\tilde{\chi}^+}(q^2)
+\Pi_{\tilde{f}}^H(q^2)+\Pi_{\tilde{f}}^D, \label{eq5} \\
\Pi_{\tilde{f}}^{g,\tilde{g}}(q^2)&=&\frac{C_f\alpha_3}{4\pi}
\left[ -A(0)-(3q^2+m_{\tilde{f}}^2)B_0(q^2,0,\tilde{f})
-2q^2B_1(q^2,0,\tilde{f})+A(\tilde{f}) \right.
\nonumber \\
&& \left. -4m_{\tilde{g}}^2B_0(q^2,\tilde{g},0)
-4q^2B_1(q^2,\tilde{g},0)\right] ,
\label{eq6} \\
\Pi_{\tilde{f}}^{\gamma Z,\tilde{\chi}^0}(q^2)
&=&\frac{Q_f^2s_W^2\alpha_2}{4\pi}
\left[ 3A(0)-(3q^2+m_{\tilde{f}}^2)B_0(q^2,0,\tilde{f})
-2q^2B_1(q^2,0,\tilde{f})+A(\tilde{f}) \right]
\nonumber \\
&& +\frac{\alpha_2}{4\pi c_W^2}(I_{3f}-s_W^2Q_f)^2
\left[ 3A(Z)-(3q^2+m_{\tilde{f}}^2)B_0(q^2,Z,\tilde{f}) \right. 
\nonumber \\
&& \left. -2q^2B_1(q^2,Z,\tilde{f})+A(\tilde{f}) \right] \nonumber \\
&& -\frac{\alpha_2}{\pi c_W^2}\sum_i|I_{3f}N_{i2}c_W+Y_fN_{i1}s_W|^2
\left[ A(0)+m_{\tilde{\chi}^0_i}^2B_0(q^2,\tilde{\chi}^0_i,0) 
\right. \nonumber \\
&& \left. +q^2B_1(q^2,\tilde{\chi}^0_i,0)\right] ,\label{eq7} \\
\Pi_{\tilde{f}}^{W,\tilde{\chi}^+}(q^2)&=&\frac{\alpha_2}{8\pi}
\left[ 3A(W)-(3q^2+m_{\tilde{f}'}^2)B_0(q^2,W,\tilde{f}')
-2q^2B_1(q^2,W,\tilde{f}')+A(\tilde{f}') \right]
\nonumber \\
&& -\frac{\alpha_2}{2\pi}\sum_j\left\{ \begin{array}{l} 
|V_{j1}|^2,\;I_{3f}=+\frac{1}{2} \\
|U_{j1}|^2,\;I_{3f}=-\frac{1}{2} \end{array} \right\} 
\left[ A(0)+m_{\tilde{\chi}^+_j}^2B_0(q^2,\tilde{\chi}^+_j,0) 
\right. \nonumber \\
&& \left. +q^2B_1(q^2,\tilde{\chi}^+_j,0)\right] ,\label{eq8} \\
\Pi_{\tilde{f}}^H(q^2)&=&\frac{\alpha_2}{4\pi c_W^2}m_Z^2(I_{3f}-Q_fs_W^2)^2
\left[ \sin^2(\alpha+\beta)B_0(q^2,h^0,\tilde{f}) \right. \nonumber \\
&& \left. +\cos^2(\alpha+\beta)B_0(q^2,H^0,\tilde{f}) \right] \nonumber \\
&&+\frac{\alpha_2}{8\pi}m_W^2
\left[ \sin^22\beta B_0(q^2,H^+,\tilde{f}')
+\cos^22\beta B_0(q^2,W,\tilde{f}') \right] . \label{eq9}
\end{eqnarray}
Here $C_q=4/3$, $C_{\ell}=0$ and $c_W\equiv\cos\theta_W$. 
$\tilde{f}'$ denotes the SU(2) partner of $\tilde{f}$. 
($U$, $V$)$_{ij}$ and $N_{ij}$ are the mixing matrices \cite{gh} for 
charginos $\tilde{\chi}^+$ and neutralinos $\tilde{\chi}^0$, respectively. 
$\sin\alpha$ is the mixing angle for neutral Higgs scalars $(h^0,H^0)$. 
The Feynman graphs corresponding to each term in eq.(\ref{eq5}) 
are shown in Fig.1. In eqs.(\ref{eq6}--\ref{eq8}), 
we have retained all $A(0)$ terms, which vanish 
after the minimal subtraction, to check the cancellation 
of quadratic divergences. 
The last term of eq.(\ref{eq5}), $\Pi_{\tilde{f}}^D$, 
denotes a special part of the scalar-loop contribution (Fig.1c) 
which is determined by the SU(2)$\times$U(1) quantum 
numbers of $\tilde{f}$. It takes a form 
\begin{equation}
\Pi_{\tilde{f}}^D=-\frac{\alpha_2}{4\pi}I_{3f}\Sigma_{W^3}
-\frac{s_W^2\alpha_2}{4\pi c_W^2}Y_f\Sigma_B. \label{eq10}
\end{equation}
$\Sigma_{W^3}$ and $\Sigma_B$ are contributions of scalar loops to 
auxiliary fields of $W^3$ and $B$, respectively, and are independent of 
the flavors of $\tilde{f}$. 
The explicit form of $\Sigma_{W^3}$ is 
\begin{eqnarray}
\Sigma_{W^3}&=&-\frac{3}{2}(A(\tilde{\nu})-A(\tilde{\ell}^-_L))
-\frac{9}{2}(A(\tilde{u}_L)-A(\tilde{d}_L))
-\frac{3}{2}(\cos^2\theta_{\tilde{t}}A(\tilde{t}_1)
+\sin^2\theta_{\tilde{t}}A(\tilde{t}_2)-A(\tilde{u}_L)) \nonumber\\
&& -\frac{\cos 2\beta}{4}(2A(H^+)-2A(W)-A(P^0)+A(Z))
-\frac{\cos 2\alpha}{4}(A(H^0)-A(h^0)). \label{eq11}
\end{eqnarray}
In eq.(\ref{eq11}), we assumed the generation independence of 
sfermion masses other than top squarks. In contrast to other terms 
(\ref{eq6}--\ref{eq9}), $\Sigma_{W^3}$ includes contributions of the 
Higgs pseudoscalar $P^0$ and top squarks with mass 
eigenstates $(\tilde{t}_1,\tilde{t}_2)$ and 
mixing angle $\theta_{\tilde{t}}$ \cite{gh}. 
The term $\Sigma_B$ does not contribute to the mass splitting (\ref{eq2}) and 
is therefore omitted here. 

The one-loop corrected mass squared splitting of the sfermion doublet 
$(\tilde{f}_1,\tilde{f}_2)$ is then 
\begin{eqnarray}
(m_{\tilde{f_1}}^2-m_{\tilde{f_2}}^2)({\rm pole})
&=&m_W^2\cos2\hat{\beta}-{\rm Re}\Pi_{\tilde{f}_1}(m_{\tilde{f}_1}^2)
+{\rm Re}\Pi_{\tilde{f}_2}(m_{\tilde{f}_2}^2) \nonumber \\
&& +{\rm Re}\Pi^{WW}_T(m_W^2)\cos2\beta
+m_W^2\sin^22\beta(\frac{\Delta v_d}{v_d}-\frac{\Delta v_u}{v_u}). 
\label{eq12}
\end{eqnarray}
In Eq.(\ref{eq12}), we eliminated the tree-level mass $\hat{m}_W$ 
by using the relation between $\hat{m}_W$ and the pole mass $m_W$: 
\begin{equation}
m_W^2=\hat{m}_W^2-{\rm Re}\Pi^{WW}_T(m_W^2)
+2m_W^2(\cos^2\beta\frac{\Delta v_d}{v_d}+\sin^2\beta\frac{\Delta v_u}{v_u}). 
\label{eq13} 
\end{equation}
The explicit form of the transverse $WW$ two-point 
function $\Pi^{WW}_T(q^2)$ in the MSSM is given 
in Refs.\cite{dhy,chankow,dabel}. 

We have to specify the renormalization condition for $\Delta v_{d,u}$ 
and $\tan\beta$ in order to study the correction to the relation (\ref{eq2}). 
In this paper, we require that $\Delta T_{d,u}^{(1)CT}$ are generated from 
the renormalization of the Higgs scalar sector and adjusted so that 
the condition 
\begin{equation}
\Delta v_d/v_d =\Delta v_u/v_u, \label{eq14}
\end{equation}
is satisfied. We then define the renormalized $\tan\beta$ by the modified 
minimal subtraction. $\tan\beta$ is then a $\overline{\rm DR}$ running 
parameter which satisfies the renormalization group equation 
\begin{equation}
\frac{d}{d\log Q}\tan\beta=-\frac{3h_t^2}{16\pi^2}\tan\beta,\;\;
h_t=\frac{gm_t}{\sqrt{2}m_W\sin\beta}, \label{eq15}
\end{equation}
for $\tan\beta\ll m_t/m_b$. 
This definition of $\tan\beta$ corresponds to ones adopted in 
refs.\cite{ay,brig,pierce} but differs from ones in 
refs.\cite{chankow,dabel}. 
Note, however, that the dependence on 
the renormalization condition for $\Delta v_{d,u}$ and $\tan\beta$ 
is common to 
both squarks and sleptons, as is seen in eq.(\ref{eq12}). 
Therefore, its choice, as well as contributions of $\Sigma_{W^3}$ and 
$\Pi^{WW}_T(m_W^2)$, does not affect the violation of the sum rule 
(\ref{eq3}) between squarks and sleptons. 

We have checked that the overall dependence of the right hand side of 
eq.(\ref{eq12}) on the renormalization scale $Q$ vanishes. 
We have also checked that the results (\ref{eq12}) is independent 
of gauge fixing parameters in general $R_{\xi}$ gauge \cite{rxi}, as long as 
eq.(\ref{eq14}) is satisfied. 

\section{Numerical Results and Discussion}
\hspace{\parindent} 
For numerical presentation of our results, we define an 
``effective $\cos2\beta$'' for a doublet 
$\tilde{F}=(\tilde{f}_1,\tilde{f}_2)_L$ as 
\begin{equation}
\cos2\beta|_{eff}^{\tilde{F}}\equiv(m_{\tilde{f}_1}^2
-m_{\tilde{f}_2}^2)/m_W^2 , \label{eq16}
\end{equation}
where all masses on the right hand side are pole masses. 
At tree-level, (16) is independent of the flavor of $\tilde{F}$ and 
identical to $\cos2\beta=(v_d^2-v_u^2)/(v_d^2+v_u^2)$. 
The ``effective $\tan\beta$'' is not useful for large $\tan\beta$ cases 
since, in such cases, eq.(\ref{eq16}) is insensitive to $\tan\beta$ and 
also $\cos2\beta|_{eff}^{\tilde{F}}<-1$ is possible, 
as we will see in this section. 

We discuss the difference between $\cos2\beta|_{eff}^{\tilde{Q}}$ 
for $\tilde{Q}=(\tilde{u}_L,\tilde{d}_L)$, 
$\cos2\beta|_{eff}^{\tilde{L}}$ for $\tilde{L}=(\tilde{\nu},\tilde{e}_L)$ 
and the $\overline{\rm DR}$ running parameter $\cos2\beta$. 
The Standard Model parameters are set as 
$m_W=80.3$~GeV, $\alpha=1/129$, $s_W^2=0.233$ and $\alpha_3=0.11$. 
The $\overline{\rm DR}$ renormalization scale is set at $m_W$. 
As for SUSY parameters, we impose the one-loop grand unification of 
gaugino masses 
$m_{\tilde{g}}/\alpha_3=M_2/\alpha_2=3M_1/(5\alpha_2\tan^2\theta_W)$ and 
generation independence of $M_{\tilde{F}}$. 
The difference between $\cos2\beta|_{eff}^{\tilde{L}}$ and 
$\cos2\beta|_{eff}^{\tilde{Q}}$ is then determined by unknown 
parameters 
($M_{\tilde{L}}$, $M_{\tilde{Q}}$, $\tan\beta$, $M_2$, $\mu$, $m_P$). 
The additional parameters ($m_t$, $m_{\tilde{t}_{1,2}}$, 
$\theta_{\tilde{t}}$), which are necessary to calculate the 
difference between $\cos2\beta|_{eff}^{\tilde{F}}$ and $\cos2\beta$, 
are chosen as $m_t=175$~GeV, 
$m_{\tilde{t}_1}^2=m_{\tilde{t}_L}^2=m_{\tilde{u}_L}^2+m_t^2$ 
and $\theta_{\tilde{t}}=0$, for simplicity. 

In Fig.2, we show the values of $\cos2\beta$, 
$\cos2\beta|_{eff}^{\tilde{L}}$ and 
$\cos2\beta|_{eff}^{\tilde{Q}}$ as functions of $\tan\beta$, for a set of 
typical values of SUSY parameters. 
The differences between $\cos2\beta|_{eff}^{\tilde{L}}$, 
$\cos2\beta|_{eff}^{\tilde{Q}}$ and $\cos2\beta$ are within $\pm0.05$. 
For the parameter choice in this figure, these differences correspond to 
about $\pm0.5$~GeV deviations of $m_{\tilde{f}_1}-m_{\tilde{f}_2}$ 
from the tree-level results. It is also shown that 
for $\tan\beta>9$, $\cos2\beta|_{eff}^{\tilde{L}}$ is below $-1$. 
In such region, obviously, we cannot define ``on-shell $\tan\beta$'' 
in terms of $m_{\tilde{f}}$ and $m_W$ by using eq.(16) and 
$\tan\beta|_{eff}^{\tilde{F}}=[(1-\cos2\beta|_{eff}^{\tilde{F}})%
/(1+\cos2\beta|_{eff}^{\tilde{F}})]^{1/2}$. 

In Fig.3, we show the difference 
$\delta c_{2\beta}\equiv\cos2\beta|_{eff}^{\tilde{L}}%
-\cos2\beta|_{eff}^{\tilde{Q}}$ for 
several values of ($M_{\tilde{L}}$, $M_{\tilde{Q}}$). The difference 
$\delta c_{2\beta}$ tends to move to negative direction 
as $m_{\tilde{L},\tilde{Q}}$ increases, and move to 
positive direction as $\tan\beta$ increases. 
We can see that the mass difference between sleptons and squarks 
is not a main cause for the difference between 
$\cos2\beta|_{eff}^{\tilde{F}}$. 
For example, its absolute 
value is larger for ($M_{\tilde{L}}=300$~GeV, $M_{\tilde{Q}}=300$~GeV) 
than for ($M_{\tilde{L}}=100$~GeV, $M_{\tilde{Q}}=300$~GeV). 
We find that the QCD ${\cal O}(\alpha_3)$ contribution (8), 
which is shown in Fig.3 for $m_{\tilde{Q}}=300$~GeV, and 
the electroweak one are of the same order. 

In Fig.4a-4c, we show $\delta c_{2\beta}$ as a function of $(M_2,\mu)$. 
The behavior of $\delta c_{2\beta}$ strongly depends on $\tan\beta$. 
For example, in the limit of $(M_2,|\mu|)\rightarrow\infty$ with fixed 
$\mu/M_2$, we find that $\delta c_{2\beta}$ logarithmically decreases 
for $\tan\beta<1.2$ but increases for $\tan\beta>1.2$. 
In Fig.4d, it is shown that the main part of the dependence of 
$\delta c_{2\beta}$ on $(M_2,\mu)$ comes from 
$\cos2\beta|_{eff}^{\tilde{Q}}$. In addition, we checked that 
the dependence of $\delta c_{2\beta}$ on $m_P$, 
which is not shown here, is smaller than 0.005 for 100~GeV$<m_P<$1~TeV. 

Finally, we consider a special case where 
$\tilde{L}$, $\tilde{Q}$ are much heavier than all other particles in the 
MSSM. This limit is theoretically interesting for the following reason: 
Suppose a possibility that the mass splitting (\ref{eq2}) and its deviation 
from the tree-level result can be described by effective 
couplings $\lambda_{\tilde{f}^*\tilde{f} H^*H}(Q=M_{\tilde{F}})$ 
in the effective 
theory where SUSY particles heavier than $M_{\tilde{F}}$ are integrated out. 
If this is the case, we can expect 
that the violation of the sum rule (\ref{eq3}) is very small in this limit, 
since, as stated in section 1, 
the difference between $\cos2\beta|_{eff}^{\tilde{L}}$ and 
$\cos2\beta|_{eff}^{\tilde{Q}}$ is generated by the violation of the 
SUSY Ward identity. 
However, this is not the case. 
Fig.5 shows the dependence on 
$M_{\tilde{Q}}=M_{\tilde{L}}\equiv m_{\tilde{F}}$ for heavy sfermions. 
We can see that in this case the 
difference $\delta c_{2\beta}$ rapidly increases as 
$m_{\tilde{F}}$ increase, contrary to naive expectation above. 
In fact, the asymptotical form of the difference $\delta c_{2\beta}$ 
in this limit is 
\begin{equation}
\delta c_{2\beta}=-\frac{4\alpha m_{\tilde{F}}}{3c_Wm_W}
+O(\alpha \log\frac{m_{\tilde{F}}}{m_W})+({\rm finite\; term}). 
\label{eq17}
\end{equation}
The first term of eq.(\ref{eq17}), proportional to $m_{\tilde{F}}$, 
emerges from 
$\tilde{f}-Z$ loop contribution in (\ref{eq7}) due to the singularity of the 
two-point functions near the threshold 
$\sqrt{q^2}\sim m_{\tilde{f}}+m_Z$. It is always larger than 
the second ``leading logarithmic'' term and dominant 
for $m_{\tilde{F}}>m_{\tilde{g}}$. 
Unfortunately, the large contribution to 
$m_{\tilde{f}_1}^2-m_{\tilde{f}_2}^2$ does not necessarily imply 
large $m_{\tilde{f}_1}-m_{\tilde{f}_2}$ which is 
experimentally observable, due to the relation 
\begin{equation}
m_{\tilde{f}_1}-m_{\tilde{f}_2}
=\frac{m_{\tilde{f}_1}^2-m_{\tilde{f}_2}^2}{m_{\tilde{f}_1}+m_{\tilde{f}_2}}
\sim \frac{m_W^2}{2m_{\tilde{F}}}\cos2\beta|_{eff}^{\tilde{F}}. \label{eq18}
\end{equation}
Therefore, in the limit $m_{\tilde{F}}\gg m_{\rm others}$, the mass 
splitting $m_{\tilde{f}_1}-m_{\tilde{f}_2}$ itself remains very small, 
about $-0.5$~GeV. Nevertheless, the large deviation from 
tree-level sum rule is theoretically very interesting. 

\section{Conclusion}
\hspace{\parindent} 
We have studied the one-loop radiative corrections to the SU(2) 
breaking mass splittings $m_{\tilde{f}_1}^2-m_{\tilde{f}_2}^2$ 
between SU(2) doublet sfermions $(\tilde{f}_1, \tilde{f}_2)$ in 
the MSSM. The analytic and numerical results of the radiative 
corrections to the mass splittings of sleptons and squarks in the 
first two generations have been presented. 
The corrections to the tree-level universal relation (\ref{eq3}) between 
sleptons and squarks is shown to be within $\pm0.05$ 
in terms of the effective $\cos2\beta|_{eff}^{\tilde{F}}$ (\ref{eq16}) 
for typical values of the SUSY parameters. We have also studied the 
dependence on SUSY parameters. The difference between 
$\cos2\beta|_{eff}^{\tilde{L}}$ and $\cos2\beta|_{eff}^{\tilde{Q}}$ 
becomes large for very heavy sfermions. 
The measurement of the violation of the relation (3) would therefore help 
us to understand the nature of the MSSM beyond the tree-level. 

\section*{Acknowledgements}
\hspace{\parindent} 
We thank M.~Nojiri for useful discussions, from which this study was 
originated. 
We also thank V.~Barger and M.~Drees for fruitful discussions and careful 
reading of the manuscript, and A.~Yamada for discussions. 
This research was supported in part by the U.S.~Department of Energy under 
Grant No.~DE-FG02-95ER40896 and in part by the University of 
Wisconsin Research Committee with funds granted by the Wisconsin 
Alumni Research Foundation.




\newpage
\vspace{20mm}
\section*{Figure Captions}
\renewcommand{\labelenumi}{Fig.\arabic{enumi}}
\begin{enumerate}

\vspace{6mm}
\item
Feynman graphs for one-loop two-point $\tilde{f}_L^*\tilde{f}_L$ functions 
in eqs.(\ref{eq5}--\ref{eq11}): 
(a) ($\Pi_{\tilde{f}}^{g,\tilde{g}}$, 
$\Pi_{\tilde{f}}^{\gamma Z,\tilde{\chi}^0}$, 
$\Pi_{\tilde{f}}^{W,\tilde{\chi}^+}$); (b) $\Pi_{\tilde{f}}^H$; 
(c) $\Pi_{\tilde{f}}^D$. Double lines denote auxiliary fields $D$ of 
corresponding gauge supermultiplets. 

\vspace{6mm}
\item
$\cos2\beta(\overline{\rm DR})$ and 
$\cos2\beta|_{eff}^{\tilde{L},\tilde{Q}}$ as functions of $\tan\beta$ for 
$M_{\tilde{L}}=M_{\tilde{Q}}=300$~GeV, $M_2=100$~GeV, $\mu=-400$~GeV and 
$m_P=300$~GeV. 

\vspace{6mm}
\item
$\delta c_{2\beta}\equiv\cos2\beta|_{eff}^{\tilde{L}}-%
\cos2\beta|_{eff}^{\tilde{Q}}$ as a function 
of $\tan\beta$ for $(M_{\tilde{L}}, M_{\tilde{Q}})({\rm GeV})=$(100, 200), 
(200, 200), (100, 300) and (300, 300). Other SUSY parameters are set as 
$M_2=100$~GeV, $\mu=-400$~GeV and $m_P=300$~GeV. 
The QCD contribution is also shown with a dashed line for 
$M_{\tilde{Q}}=300$~GeV. 

\vspace{6mm}
\item
$\delta c_{2\beta}$ in the 
$(M_2, \mu)$ plane for $\tan\beta=$1.1 (a), 2 (b), 10 (c) and 
($\cos2\beta|_{eff}^{\tilde{L}}$ (thin dashed lines), 
$\cos2\beta|_{eff}^{\tilde{Q}}$ (solid lines)) for 
$\tan\beta=2$ (d). Other SUSY 
parameters are set as $M_{\tilde{L}}=M_{\tilde{Q}}=m_P=300$~GeV. 
The regions below the thick dashed lines, where either 
$2m_{\tilde{\chi}^+_1}<m_Z$ 
or $m_{\tilde{\chi}^0_1}+m_{\tilde{\chi}^0_2}<m_Z$ holds, are 
excluded by LEP-I constraints. 

\vspace{6mm}
\item
$\cos2\beta|_{eff}^{\tilde{L},\tilde{Q}}$ as functions of 
$m_{\tilde{F}}\equiv M_{\tilde{L}}=M_{\tilde{Q}}$. 
Other SUSY parameters are set as $\tan\beta=2$ ($\cos2\beta=-0.6$), 
$M_2=100$~GeV, $\mu=-400$~GeV and $m_P=300$~GeV. 

\end{enumerate}

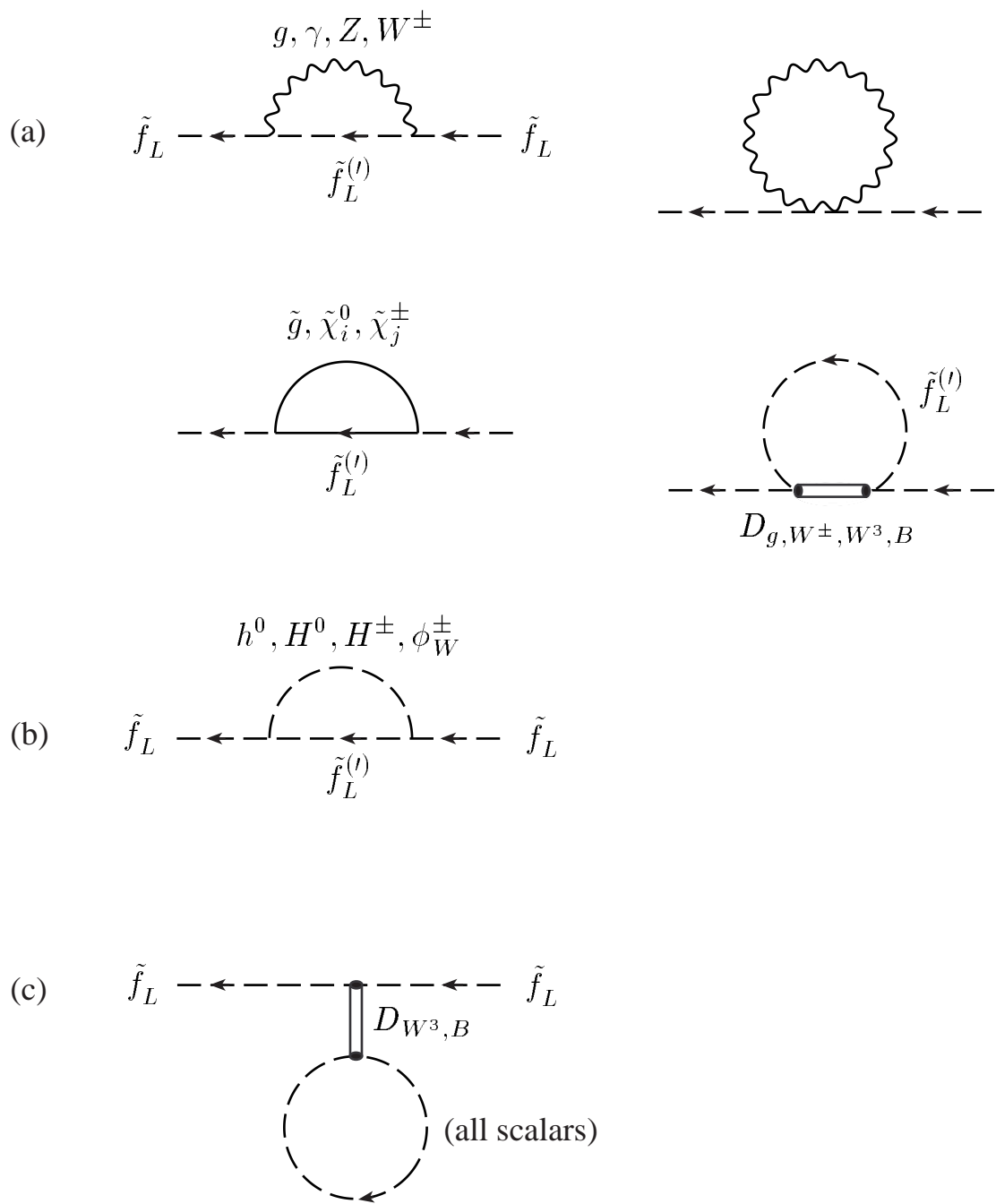
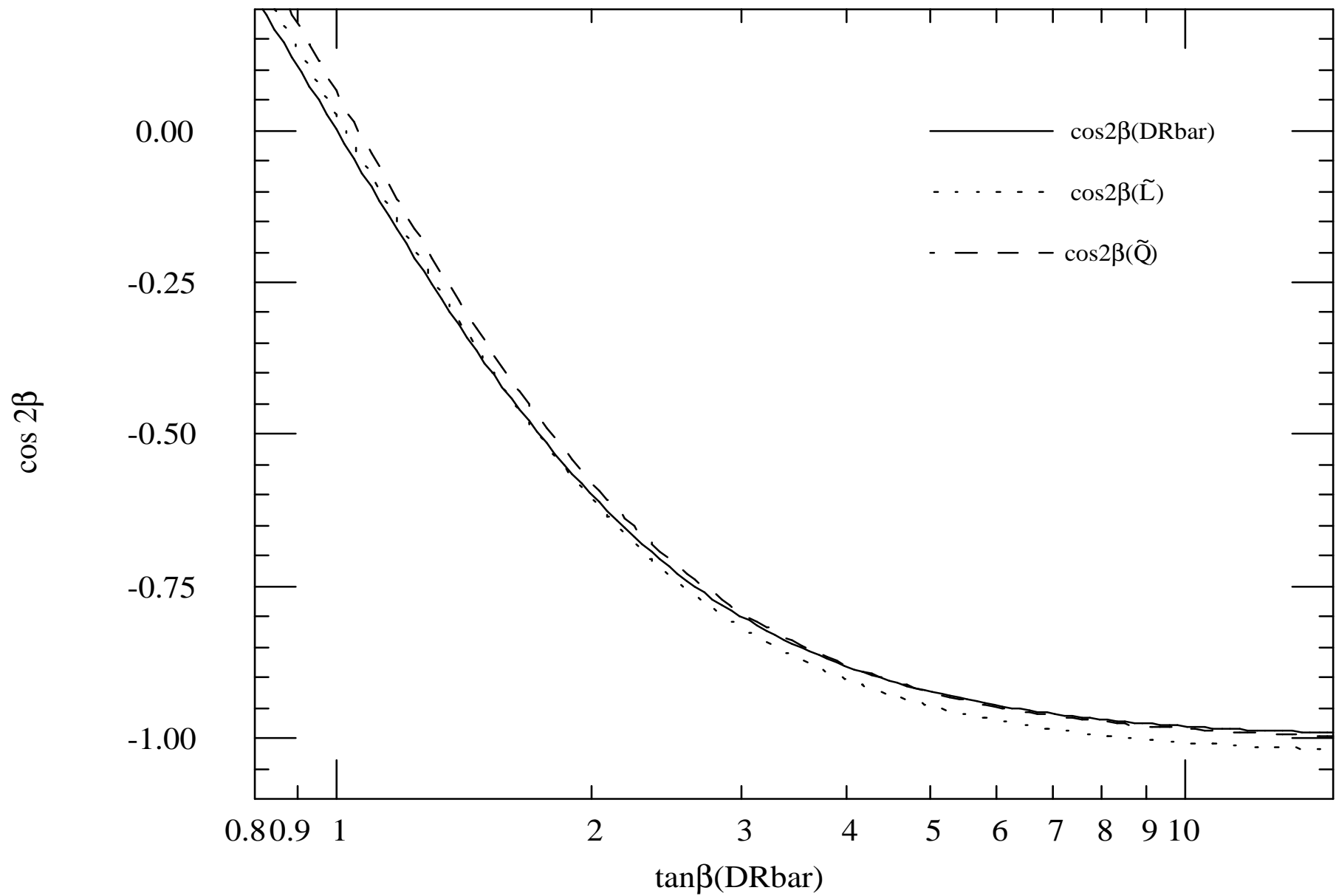
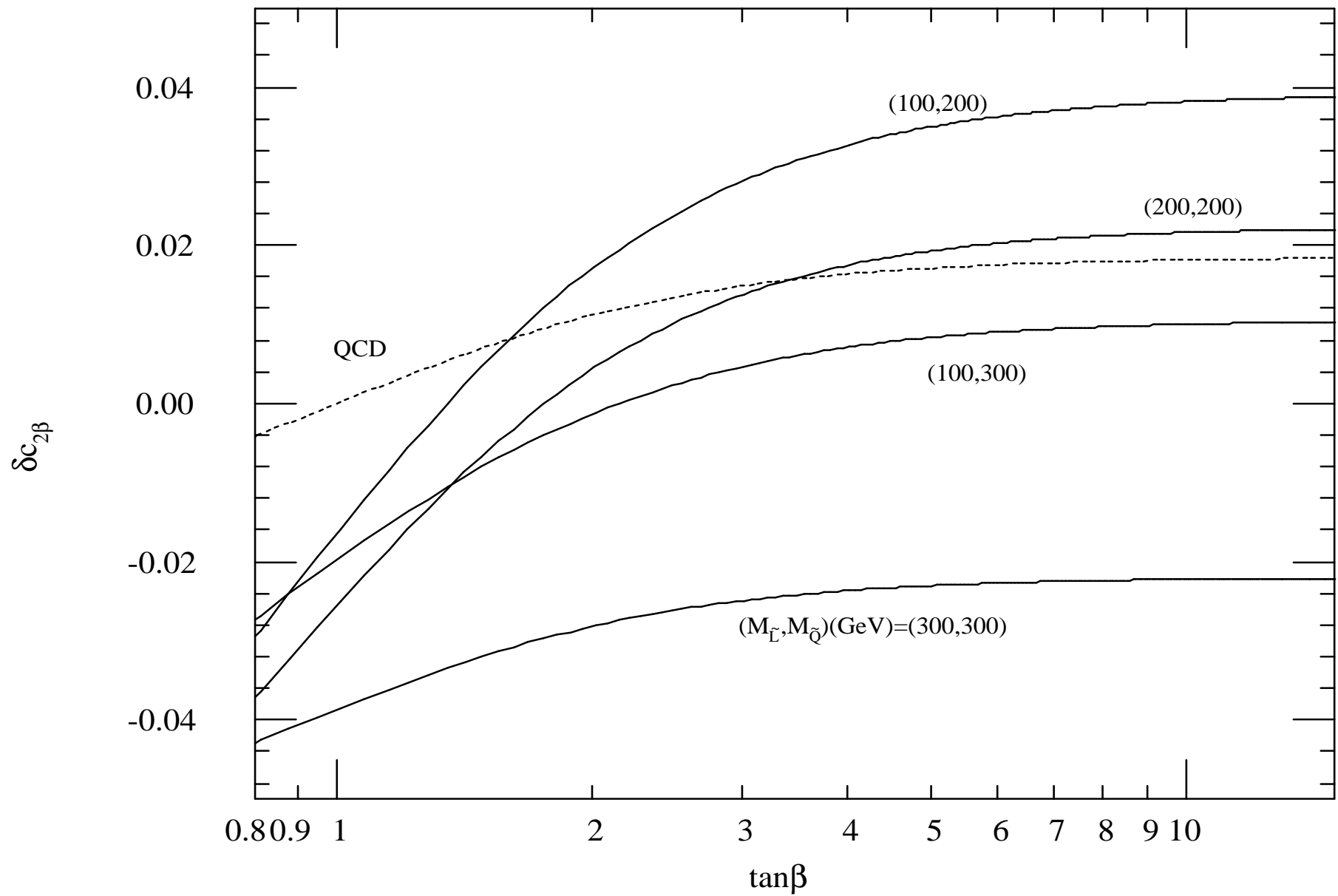
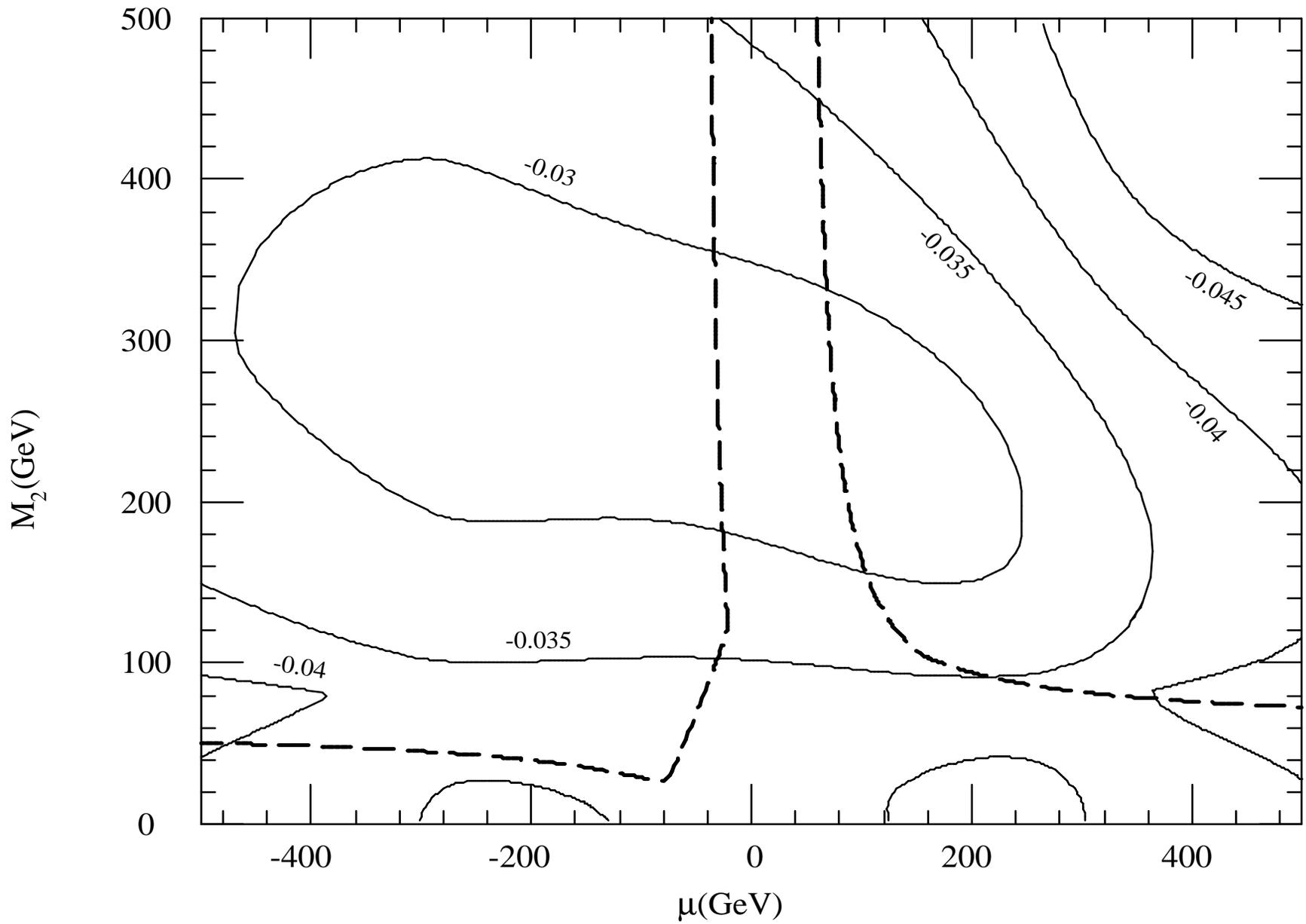
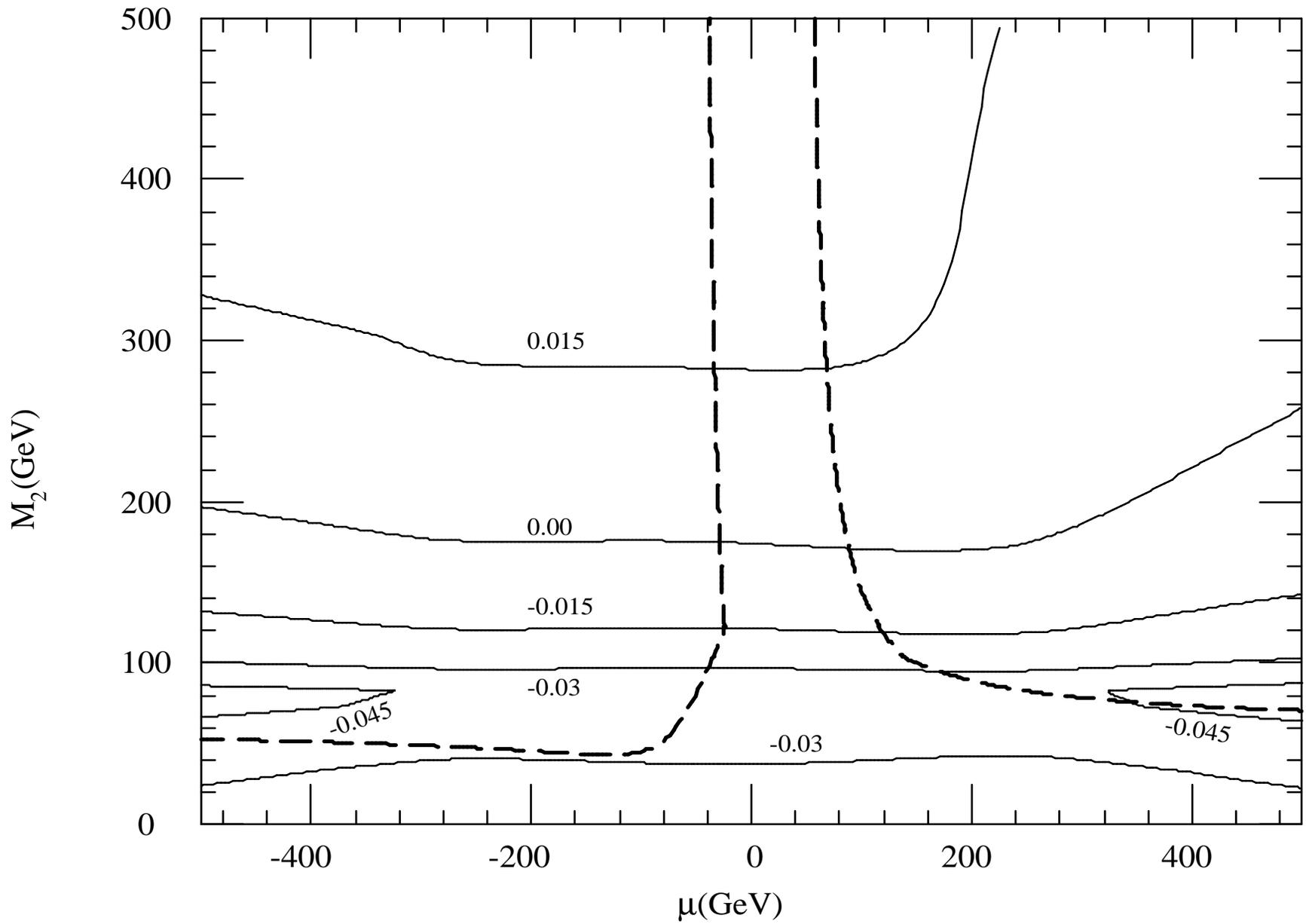
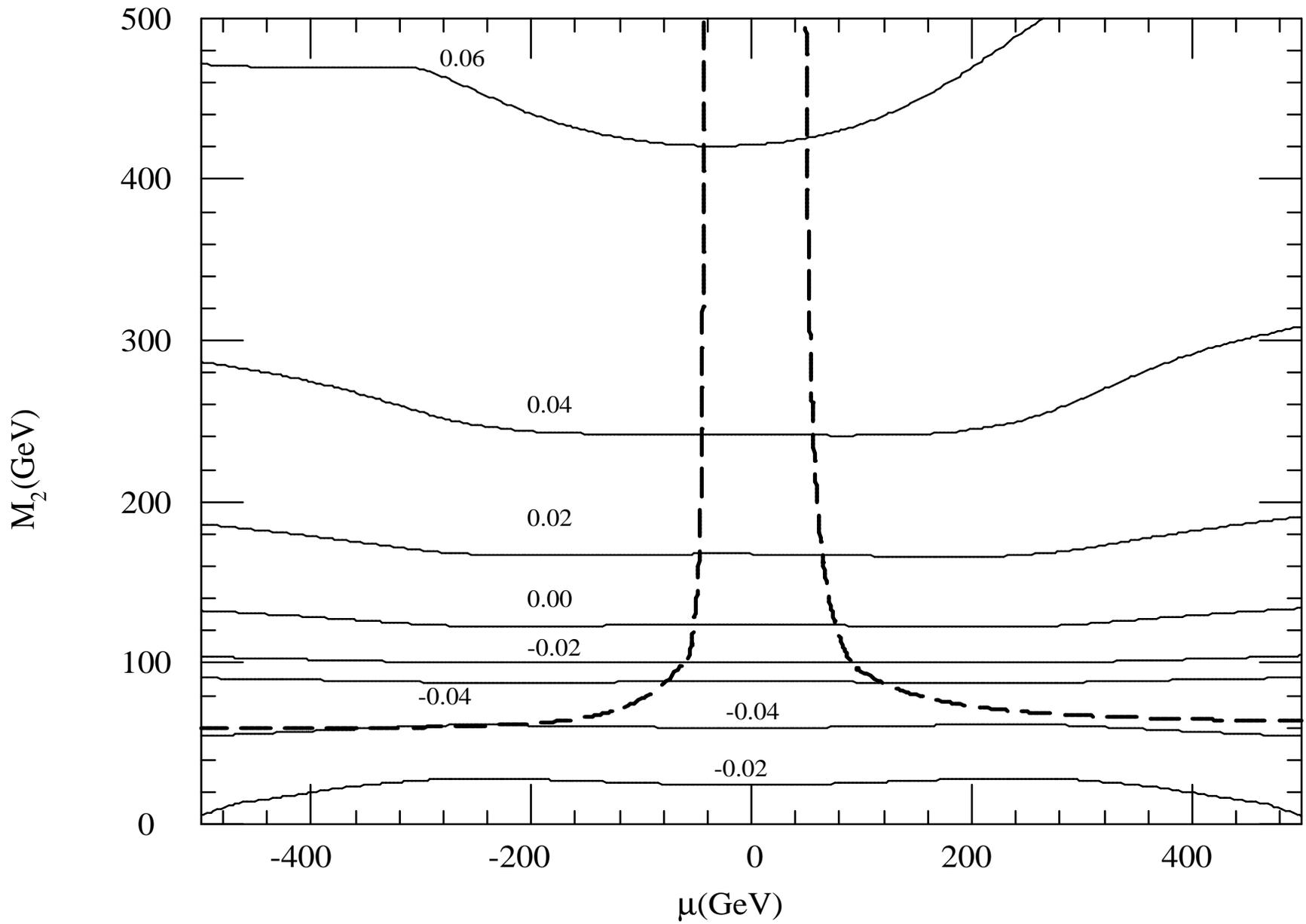
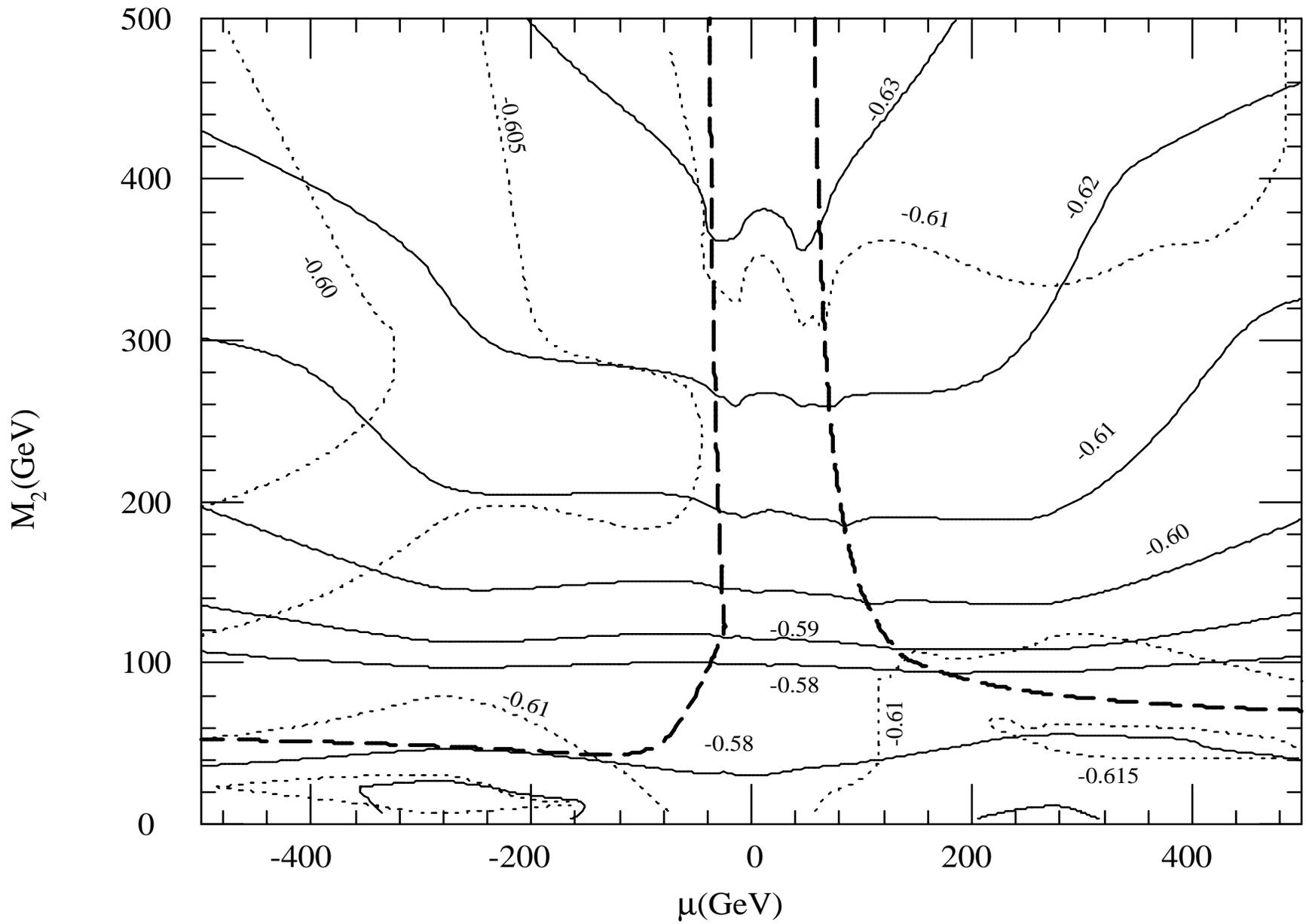
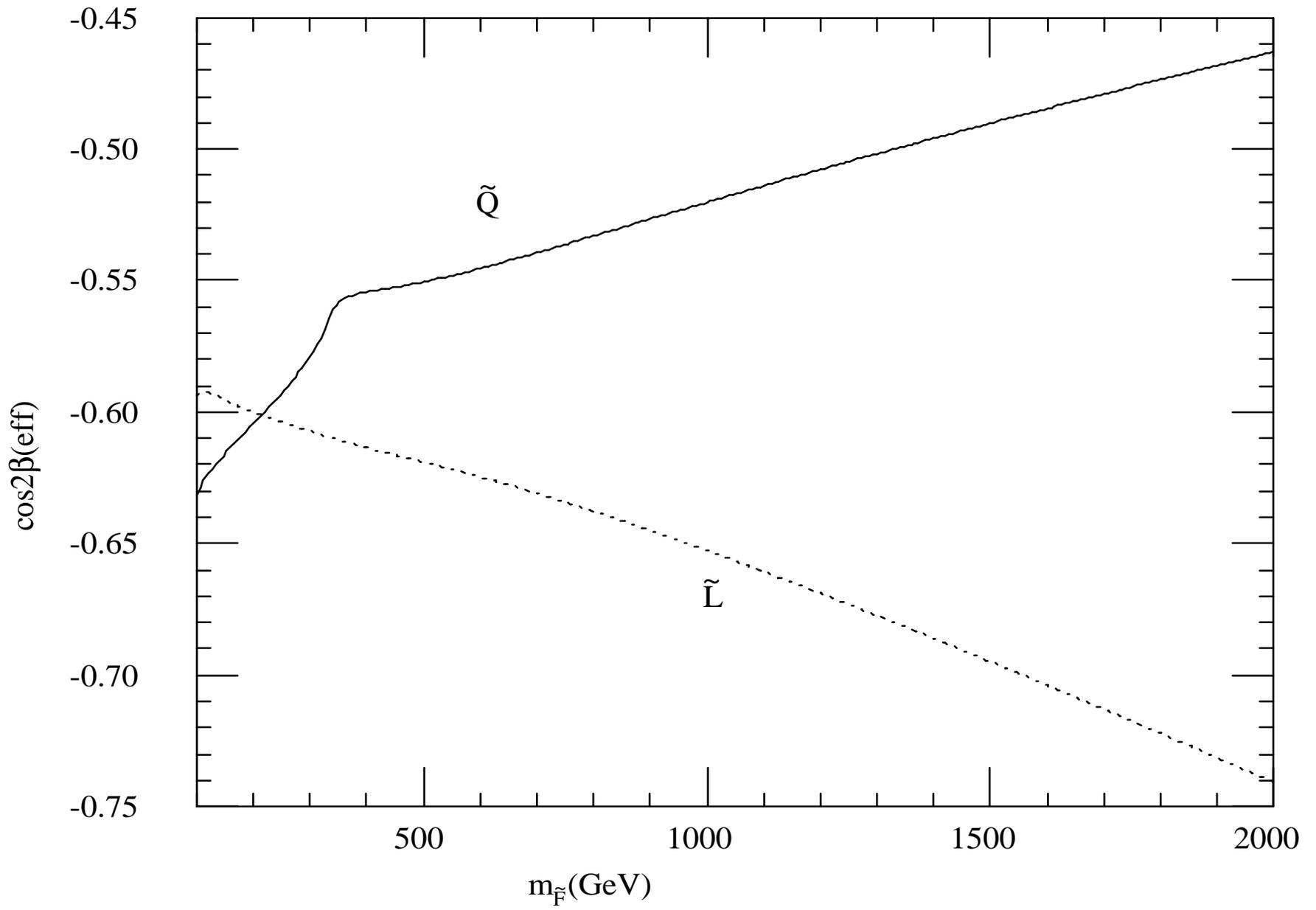
\end{document}